\documentclass[aps,prl,twocolumn]{revtex4}

\begin{document}

\title{Comment on ``Stable Quantum Computation of Unstable Classical Chaos''}

\author{Christof Zalka}
\email{zalka@uwaterloo.ca}
\affiliation{Department of Combinatorics and Optimization, University of
Waterloo, Waterloo, Ontario, Canada N2L 3G1}

\maketitle
In a recent letter, Georgeot and Shepelyansky \cite{gs2} claim to have
shown that there is a big advantage in using a quantum computer to
simulate and study classical chaotic systems, over using a
conventional (classical) computer. (Actually, the authors made this
claim already in the second part of an earlier letter \cite{gs1}.)

Most of the paper \cite{gs2} is devoted to showing that the quantum
simulation is much less affected by errors than the classical
one. This is based on a clearly inappropriate comparison of the two
cases. The authors consider the computationally simple, but chaotic,
``Arnold cat map'', which is an area-preserving bijective map of a
two-dimensional torus onto itself. They also consider a discretized
version of this map which can be simulated exactly on a digital
computer. The authors then propose to compute this discretized map on
a quantum computer on many possible initial values simultaneously,
utilizing ``quantum parallelism''.

Strangely they then compare this quantum computation of the
discretized map to a classical simulation of the full continuous
map. Because they imagine the classical simulation to be done with
fixed-precision arithmetic, the chaotic dynamics will quickly amplify
rounding errors, so that after a few iterations of the map, the
results will be totally wrong. It is then clear that the quantum
simulation fares better, even when considering noise acting on the
qubits, which the authors do. For this comparison, see e.g. figure 1
(left classical, right quantum) or the last paragraph of the paper.

Thus, in effect, the authors reach their conclusion in favor of the
quantum simulation by demanding more from the classical than from the
quantum simulation, specifically, that it should do an accurate
simulation for any real initial values. Actually, the issue of
rounding errors is not specific to classical or quantum computation,
but is typical for {\it digital} computation. Note that the authors
consider the usual kind of quantum computer consisting of qubits which
{\it is} digital, namely, binary.

Apart from the discussion of errors, the authors make an
unsubstantiated claim just before the last paragraph. They say that
their quantum simulation lets one obtain ``global quantities
inaccessible by classical computation''. They propose to model the
initial superposition of input values according to some (classical)
probability density distribution (how, exactly, they do not say). They
then want to extract information of interest about the final density,
for example, by applying a quantum Fourier Transformation to the final
superposition before observing the quantum computer.

Although this may allow the extraction of some information about the
power spectrum of the final distribution, it is by no means clear that
this could not be done just as efficiently on a classical computer. In
particular, note that for the kind of quantum simulation described
here, we need a {\it reversible} discrete map. This would, for
example, allow one to evolve a few closely spaced final values
backwards (on a classical computer), thereby permitting one to obtain
information about the ``fine structure'' of a final density.

At any rate, it is an extraordinary claim, worthy of careful
justification, to have found a new application for which a quantum
computer provides exponential advantage over the known classical
algorithms. This is especially true for ``openly given'' problems,
thus without ``black boxes'' or limitations on communication etc.  So
far not much more than Shor's algorithm(s) and, not surprisingly, the
simulation of quantum systems achieved that.


\begin{thebibliography}{10}

\bibitem{gs2} B. Georgeot and D.L. Shepelyansky,
Phys. ~Rev. ~Lett. {\bf 86} (23),
 5393 (2001) \qquad (also quant-ph/0101004)

\bibitem{gs1} B. Georgeot and D.L. Shepelyansky,
Phys. ~Rev. ~Lett. {\bf 86} (13),
 2890 (2001) \qquad (also quant-ph/0010005)

\end{thebibliography}
\end{document}